\documentclass[aps,prl,reprint,groupedaddress]{revtex4-1}
\usepackage{graphicx}
\newcommand{\mnras}{Mon. Not. R. Astron. Soc.}
\newcommand{\aap}{Astron. Astrophys.}

\newcommand{\araa}{Ann. Rev. Astron. Astrophys.}
\newcommand{\apjl}{Astrophys. J.}
\newcommand{\pasj}{Publ. Astron. Soc. Jpn.}
\newcommand{\jcap}{J. Cosmol. Astropart. Phys.} 
\newcommand{\ssr}{Space Science Reviews} 
\begin{document}

\title{Hadronic origin of multi-TeV gamma rays and neutrinos from
low-luminosity active galactic nuclei: implications of past activities
of the Galactic center}

\author{Yutaka Fujita}
\affiliation{Department of Earth and Space Science, Graduate School of
 Science, Osaka University, Toyonaka, Osaka 560-0043, Japan}

\author{Shigeo S. Kimura} \affiliation{Frontier Research Institute for
Interdisciplinary Sciences, Tohoku University, Sendai, 980-8578, Japan}
\affiliation{Astronomical Institute, Tohoku University, Sendai 980-8578,
Japan}

\author{Kohta Murase} \affiliation{Center for Particle and Gravitational
Astrophysics; Department of Physics; Department of Astronomy \&
Astrophysics, The Pennsylvania State University, University Park,
Pennsylvania, 16802, USA} 
\affiliation{Institute for Advanced Study,
Princeton, New Jersey 08540, USA}

\date{\today}

\begin{abstract}
Radiatively inefficient accretion flows (RIAFs) in low-luminosity active
galactic nuclei (LLAGNs) have been suggested as cosmic-ray and neutrino
sources that may largely contribute to the observed diffuse neutrino
intensity. We show that this scenario naturally predicts hadronic
multi-TeV gamma-ray excesses around Galactic centers.  The protons
accelerated in the RIAF in Sagittarius A$^*$ (Sgr A$^*$) escape and
interact with dense molecular gas surrounding Sgr A$^*$, which is known
as the central molecular zone (CMZ), and produce gamma rays as well as
neutrinos. Based on a theoretical model that is compatible with the
IceCube data, we calculate gamma-ray spectra of the CMZ and find that
the gamma rays with $\gtrsim 1$~TeV may have already been detected with
the High Energy Stereoscopic System, if Sgr A$^*$ was more active in the
past than it is today as indicated by various observations. Our model
predicts that neutrinos should come from the CMZ with a spectrum similar
to the gamma-ray spectrum. We also show that such a gamma-ray excess is
expected for some nearby galaxies hosting LLAGNs.
\end{abstract}


\maketitle


\section{I. Introduction}

Recently, the IceCube Collaboration reported the discovery of
extraterrestrial neutrinos
\cite{2013PhRvL.111b1103A,2014PhRvL.113j1101A,2015PhRvD..91b2001A}. The
origin of the neutrinos is a matter of a debate
(\cite{2006JCAP...05..003L,2008ApJ...689L.105M,2013PhRvD..88l1301M,2013PhRvL.111l1102M,2013PhRvD..88d7301S,2013arXiv1311.0287K,2014PhRvD..90b3010A,2014JCAP...09..043T,2014PhRvD..90l3012Y,2014ApJ...793..131C,2013ApJ...774...74F,2014PhRvD..89h3004L,2014JHEAp...3...29D,2014PhRvD..90b3007M,2014ApJ...790L..14K,2015ApJ...806..159K}
for reviews, see \cite{2014NuPhS.256..241M,2014arXiv1410.3680M}). The
data so far are compatible with an isotropic distribution, which
suggests that neutrinos are of extragalactic origin.  Diffuse gamma-ray
data also support this idea
\cite{2013PhRvD..88l1301M,2014PhRvD..90b3010A}. However, the sources
have not been identified because of poor angular information and
statistics for the neutrinos. One way of improving this situation may be
detection of counterparts through electromagnetic waves.

Low-luminosity active galactic nuclei (LLAGNs) are a candidate for the
source of the neutrinos. The LLAGNs are expected to have radiatively
inefficient accretion flows (RIAFs), which are realized when the mass
accretion rate into the supermassive black hole (SMBH) is relatively
small ($\dot{M}/\dot{M}_{\rm Edd}\lesssim 0.01$--0.1), where
$\dot{M}_{\rm Edd}=L_{\rm Edd}/c^2$ is the Eddington accretion rate, and
$L_{\rm Edd}$ is the Eddington luminosity \cite{2014ARA&A..52..529Y}.
In the tenuous and turbulent plasma in the RIAFs, cosmic ray (CR)
protons may be accelerated via stochastic acceleration or magnetic
reconnection \cite{2015ApJ...806..159K}. These CR protons interact with
other nucleons ($pp$ interaction) and photons ($p\gamma$ interaction) in
the flow and generate neutrinos. Although the production rate of the
neutrinos per an LLAGN is not large compared with other more energetic
sources such as quasistellar objects (QSOs), the abundance of LLAGNs
can reproduce observed neutrino flux on the Earth
\cite{2015ApJ...806..159K}. Even if it is difficult to resolve LLAGNs as
point neutrino sources, gamma rays from them could be used to test this
model. In particular, the gamma rays that are a byproduct of the
$pp$ interactions have energies comparable to those of the neutrinos,
which means that the gamma-ray spectrum should reflect the neutrino
spectrum unless the gamma rays are not absorbed.

Sagittarius A$^*$ (Sgr~A$^*$) is the SMBH at the center of the Galaxy
and it is known as an LLAGN. The current mass accretion rate of
Sgr~A$^*$ is very small and the accretion flow is thought to be a RIAF
\cite{2003ApJ...598..301Y}. The current production rate of CR protons in
the RIAF is expected to be small because of the small accretion rate
\cite{2014ApJ...791..100K}. Thus, the gamma-ray luminosity of the RIAF
in the TeV band is also expected to be small because of inefficient pion
production \cite{2015ApJ...806..159K}. However, it has been indicated
that Sgr~A$^*$ was much more active in the past
\cite{1996PASJ...48..249K,2000ApJ...534..283M,2006PASJ...58..965T,2013PASJ...65...33R}. During
those activities, a large amount of protons could have been accelerated
and escaped from the RIAF.  Moreover, observations have revealed that
there is a huge amount of molecular gas surrounding Sgr~A$^*$. This gas
concentration is known as the central molecular zone (CMZ) with the size
of $R_{\rm CMZ}\sim$100~pc and the mass of $M_{\rm CMZ}\sim 10^7\:
M_\odot$ \cite{1996ARA&A..34..645M}. Strong turbulence and magnetic
fields in the CMZ may delay the diffusion of the CR protons that have
plunged into the CMZ, and those protons may stay in the CMZ for a long
time. In this paper, we calculate the diffusion of the protons in the
CMZ that have accelerated and escaped from the RIAF in Sgr~A$^*$. We
estimate gamma-ray and neutrino emissions created through $pp$
interactions between the CR protons and protons in the CMZ. We show that
TeV gamma rays from the CMZ around Sgr~A$^*$ and nearby LLAGNs observed
with the High Energy Stereoscopic System (HESS) may be produced by this
mechanism. Previous studies have shown that the gamma-ray excess
observed in those objects at $\gtrsim 1$~TeV cannot be explained by
one-zone leptonic models
\cite{2012ApJ...748...34K,2014A&A...562A..12P,2014ApJ...790...86Y}. Our
model can resolve this issue, although there are also studies trying to
explain the HESS observations by $pp$ interactions in the CMZ by
different approaches
\cite{2011MNRAS.410L..23M,2011PASJ...63L..63A,2014ApJ...790...86Y,2007ApJ...657L..13B,2009APh....31...13D}.

\section{II. COSMIC-RAY PROTON ACCELERATION IN RADIATIVELY INEFFICIENT 
ACCRETION FLOWS OF SGR~A$^*$}

In our model, protons are accelerated in the RIAF of Sgr~A$^*$. Since
the acceleration is confined in a small region on a scale of a few tens
of the Schwarzschild radius of the SMBH, we consider the acceleration
based on a one-zone model as previous studies
\cite{2012ApJ...746..164M,2015ApJ...806..159K,2015MNRAS.449..551K}. According
to the model of Ref.~\cite{2015ApJ...806..159K}, the typical energy of
the accelerated protons is determined by the balance between the
acceleration time of the protons in a RIAF ($t_{\rm acc,R}$) and their
escape time from the RIAF. The escape time is comparable to the
diffusion time of the protons in the RIAF ($t_{\rm diff,R}$). Thus, the
Lorentz factor corresponding to the typical energy is obtained by
solving the equation of $t_{\rm acc,R}=t_{\rm diff,R}$ and the result is
\begin{eqnarray}
\label{eq:geq}
 \frac{E_{p,\rm eq}}{m_p c^2} &\sim& 1.4\times 10^5 
\left(\frac{\dot{m}}{0.01}\right)^{1/2}
\left(\frac{M_{\rm BH}}{1\times 10^7\: M_\odot}\right)^{1/2} \nonumber\\
&\times &\left(\frac{\alpha}{0.1}\right)^{1/2}
\left(\frac{\zeta}{0.1}\right)^3
\left(\frac{\beta}{3}\right)^{-2}
\left(\frac{R_{\rm acc}}{10\: R_S}\right)^{-7/4}\:,
\end{eqnarray}
where $m_p$ is the proton mass, $\dot{m}$ is the normalized accretion
rate $\dot{m}=\dot{M}/\dot{M}_{\rm Edd}$, $\alpha$ is the alpha
parameter of the accretion flow \cite{1973A&A....24..337S}, $\zeta$ is
the ratio of the strength of turbulent fields to that of the
nonturbulent fields, $\beta$ is the plasma beta parameter, $R_{\rm
acc}$ is the typical radius where particles are accelerated, and $R_S$
is the Schwarzschild radius of the black hole
\cite{2015ApJ...806..159K}. For the parameters of the RIAF, we take
$\alpha=0.1$, $\zeta=0.05$, $\beta=3$, and $R_{\rm acc}=10\: R_S$ as
fiducial parameters, because the neutrino flux at $\sim 10$--100~TeV
obtained with IceCube is well reproduced by them
\cite{2015ApJ...806..159K}.

We assume that the luminosity of the protons accelerated in the RIAF is
$L_{p,\rm tot}=\eta_{\rm cr} \dot{M}c^2$, where $\eta_{\rm cr}$ is the
parameter and we take $\eta_{\rm cr}=0.015$ as the fiducial value
following Ref.~\cite{2015ApJ...806..159K}. When only stochastic
acceleration is effective, the production rate of protons in the
momentum range $p$ to $p+dp$ is
\begin{equation}
\label{eq:sp}
 \dot{N}(x)dx \propto x^{(7-3q)/2}K_{(b-1)/2}(x)dx\:,
\end{equation}
where $x=p/p_{\rm cut}$, $K_\nu$ is the Bessel function, and $b=3/(2-q)$
\cite{2006ApJ...647..539B}. The power-law index of turbulence
responsible for the acceleration is assumed to be $q=5/3$ (Kolmogorov
type). The cutoff momentum is defined as $p_{\rm
cut}=(2-q)^{1/(2-q)}p_{\rm eq}=p_{\rm eq}/27$, where $p_{\rm
eq}=E_{p,{\rm eq}}/c$ \cite{2015ApJ...806..159K,2006ApJ...647..539B}. We
determine the normalization of Eq.~(\ref{eq:sp}) so that the total power
of the protons is $L_{p,\rm tot}$.

\section{III. Diffusion of Protons in the CMZ}

Protons accelerated in Sgr~A$^*$ leave the acceleration site (RIAF) and
disperse into the interstellar space. Some of them would enter the CMZ
surrounding Sgr~A$^*$. We solve a diffusion-convection equation for the
CR protons in the CMZ. For the sake of simplicity, we solve a
spherically symmetric equation:
\begin{equation}
\label{eq:diff}
 \frac{\partial f}{\partial t}
= \frac{1}{r^2}\frac{\partial}{\partial r}\left(r^2 \kappa\frac{\partial
					       f}{\partial r}\right)
- u\frac{\partial f}{\partial r}
+ \frac{1}{3 r^2}\left[\frac{\partial}{\partial r}(r^2 u)\right]
p\frac{\partial
f}{\partial p} + Q\:,
\end{equation}
where $f=f(t,r,p)$ is the distribution function, $r$ is the distance
from the Galactic center, $p$ is the momentum of particles, $\kappa$ is
the diffusion coefficient, $u$ is the velocity of the background gas,
and $Q$ is the source term for the particles (Sgr~A$^*$). We assume that
$u=0$, because we are interested in the CRs inside the CMZ, which is too
heavy to be moved by possible outflows from Sgr~A$^*$. We do not
consider CRs carried by the outflows without entering into the CMZ. We
assume that the CMZ is uniform and its dense gas occupies at $r<R_{\rm
CMZ}$.

The actual CMZ has a disclike structure and does not entirely cover
Sgr~A$^*$ \cite{1996ARA&A..34..645M}. Thus, we expect that most of the
CR protons do not plunge into the CMZ, and we assume that only a
fraction $\lambda$ of the protons accelerated in the RIAF are injected
into the CMZ. Thus, the source term in Eq.~(\ref{eq:diff}) is written as
$\int 4\pi c p^3 Q dp =\lambda L_{p,\rm tot} = \lambda\eta_{\rm cr}
\dot{M}c^2$. Since the size of the CMZ ($R_{\rm CMZ}\sim 100$~pc) is
much larger than that of the RIAF, we treat $Q$ as a point source.

We assume that the diffusion coefficient of CRs outside the RIAF is
given by
\begin{equation}
\label{eq:diffc}
 \kappa = 10^{28} \left(\frac{E_p}{10\rm\; GeV}\right)^{0.5}
\left(\frac{B}{3\rm\: \mu G}\right)^{-0.5} \rm\: cm^2 s^{-1}\:,
\end{equation}
where $E_p$ is the particle energy. This coefficient is for the ordinary
interstellar space in the Galactic disc \cite{2009MNRAS.396.1629G}.  We
only consider resonant scattering, which is valid at sufficiently low
energies. Although the diffusion coefficient in the CMZ is not known, we
apply Eq.~(\ref{eq:diffc}) to stronger magnetic field cases.  If there
are strong magnetic fields ($B\sim$~mG) in the CMZ
\cite{1996ARA&A..34..645M}, the coefficient is much smaller than that in
the intercloud space around the Galactic center ($B\sim \rm\: 10 \mu G$)
\cite{2009A&A...505.1183F}. In fact, it has been indicated that the
diffusion coefficient in molecular gas around supernova remnants is much
smaller than the ordinary value \cite{2009ApJ...707L.179F}. From now on,
we fix magnetic fields at $B=1$~mG for $r<R_{\rm CMZ}$ and $10\rm\: \mu
G$ for $r>R_{\rm CMZ}$. We do not consider stochastic acceleration in
the CMZ. This is because the diffusion coefficient we assumed in
Eq.~(\ref{eq:diffc}) is too large for effective particle
acceleration. In fact, previous studies have shown that the diffusion
coefficient must be as small as that for the Bohm diffusion for
particles to be accelerated up to $\gtrsim $~TeV
\cite{2011PASJ...63L..63A,2012ApJ...750...21F}. It is not certain
whether such a small coefficient is realized by turbulence in and around
the CMZ.  We do not include cooling of the protons in
Eq.~(\ref{eq:diff}), because the cooling time is larger than the
diffusion time estimated based on Eq.~(\ref{eq:diffc}) (see later).

CR protons interact with protons in the CMZ. For a given distribution
function $f$, we calculate the production rate of gamma-ray photons
using the code provided by \cite{2008ApJ...674..278K} and the formula
provided by \cite{2006PhRvD..74c4018K} for $E_p<1$ and $E_p>1$~TeV,
respectively. The results are not sensitive to the boundary energy
(1~TeV). We also calculate the neutrino production rate at the same
time. We consider the attenuation of very high energy gamma rays by pair
production on the Galactic interstellar radiation field using the
results shown in Fig.~3 of Ref.~\cite{2006ApJ...640L.155M}. However, the
following results are not much affected by the attenuation. The energy
density of interstellar radiation field ($\sim 10\rm\; eV\: cm^{-3}$) is
much smaller than that of the assumed magnetic field ($\sim$~mG)
\cite{2005ICRC....4...77P}. Thus, the gamma-ray emission via inverse
Compton scattering by secondary electrons can be ignored.

\section{IV. gamma rays from the CMZ around Sgr~A$^*$}

Observations showed that the radius of the CMZ at the Galactic center is
$R_{\rm CMZ,obs}=200$~pc, the thickness is $H_{\rm CMZ,obs}=75$~pc, and
the mass is $M_{\rm CMZ}=2\times 10^7\: M_\odot$
\cite{2000ApJ...545L.121P,2014ApJ...790..109M}.  Thus, the average
density is $\rho_{\rm CMZ}=M_{\rm CMZ}/(\pi R_{\rm CMZ,obs}^2H_{\rm
CMZ,obs})=1.4\times 10^{-22}\rm\: g\: cm^{-3}$. Since
Eq.~(\ref{eq:diff}) assumes spherical symmetry, we define an effective
radius
\begin{equation}
 R_{\rm CMZ}\equiv 
\left(\frac{3\: M_{\rm CMZ}}{4\:\pi\rho_{\rm CMZ}}\right)^{1/3}
=130\:\rm pc\:,
\end{equation}
and we use this as the radius of the CMZ in the following calculations.
The covering factor of the CMZ for $r<R_{\rm CMZ,obs}$ is $f_{\rm
CMZ}=\pi R_{\rm CMZ,obs}^2 H_{\rm CMZ,obs}/(4\pi R_{\rm
CMZ,obs}^3/3)=0.5$, The fraction of CR protons that enter into the CMZ,
$\lambda$, may be comparable to $f_{\rm CMZ}$. However, if the protons
are not spherically emitted and they are, for example, carried by strong
outflows perpendicular to the disclike CMZ, the fraction may be much
smaller. Moreover, the inner edge of the CMZ may not have contact with
Sgr~A$^*$. Thus, we assume that $\lambda\leq 0.5$, and $\lambda\ll 0.5$
is very likely. Observations have shown that the mass of the SMBH is
$M_{\rm BH}=4.3\times 10^6\: M_\odot$ \citep{2009ApJ...692.1075G}, and
the current mass accretion rate is $\dot{M}= 4\times 10^{-8}\: M_\odot
\rm\: yr^{-1}$ \cite{2003ApJ...598..301Y}. Since the Eddington
luminosity is given by $L_{\rm Edd}=1.26\times 10^{38}\: (M_{\rm
BH}/M_\odot) \:\rm erg\: s^{-1}$, the Eddington accretion rate is
$\dot{M}_{\rm Edd}=9.6\times 10^{-3}\: M_\odot \rm\: yr^{-1}$. Thus, the
normalized accretion rate is written as $\dot{m}=4.2\times 10^{-6}$. For
these and the fiducial parameters, we obtain $E_{p,\rm eq}=0.2\rm\: TeV$
from Eq.~(\ref{eq:geq}). We solve Eq.~(\ref{eq:diff}) from $t=0$ to
$10^7$~yr. There are no CRs at $t=0$. The distribution of CRs has
achieved a steady state at the end of the simulation because the
diffusion time of CRs is only $t_{\rm diff,C}=R_{\rm
CMZ}^2/(6\kappa)\sim 1.6\times 10^5$~yr at $E_p\sim 1$~TeV. The
following results are those at $t_0=10^7$~yr. Since this time scale is
much larger than $t_{\rm diff,C}$, the energy spectrum of the CR protons
is almost uniform in the CMZ
\cite{1995PhRvD..52.3265A,2009ApJ...707L.179F}. As long as the spectrum
of the injected CRs does not vary significantly, our model does not
expect substantial variation in the gamma-ray spectrum across the CMZ
(see later), which is consistent with observations. The distance to
Sgr~A$^*$ and the CMZ is assumed to be 8.5~kpc.

The inverse of the cooling time of CR protons due to $pp$ interactions
is
\begin{equation}
 t_{pp}^{-1} \sim n_{\rm CMZ}\sigma_{pp}c K_{pp}\:,
\end{equation}
where $n_{\rm CMZ}=\rho_{\rm CMZ}/m_p$ and $K_{pp} (\sim 0.5)$ is the
proton inelasticity of the process. The total cross section of the
process is given by
\begin{equation}
 \sigma_{pp} = (34.3 + 1.88 L + 0.25 L^2)
\left[1-\left(\frac{E_{\rm th}}{E_p}\right)^4\right]^2\rm\: mb\:,
\end{equation}
where $E_{\rm th}=1.22$~GeV is the threshold energy of production of
$\pi^0$ mesons and $L=\ln(E_p/1\rm\: TeV)$ \cite{2006PhRvD..74c4018K}.
Thus, the cooling time is $t_{pp}\sim 7\times 10^5$~yr at $E_p\sim
1$~TeV, which is larger than $t_{\rm diff,C}$. Since $t_{pp}>t_{\rm
diff,C}$ is satisfied in the energy of interest ($E_p\gtrsim 1$~TeV), we
do not need to include the cooling effect in Eq.~(\ref{eq:diff}).
 
\begin{figure}
\includegraphics[width=90mm]{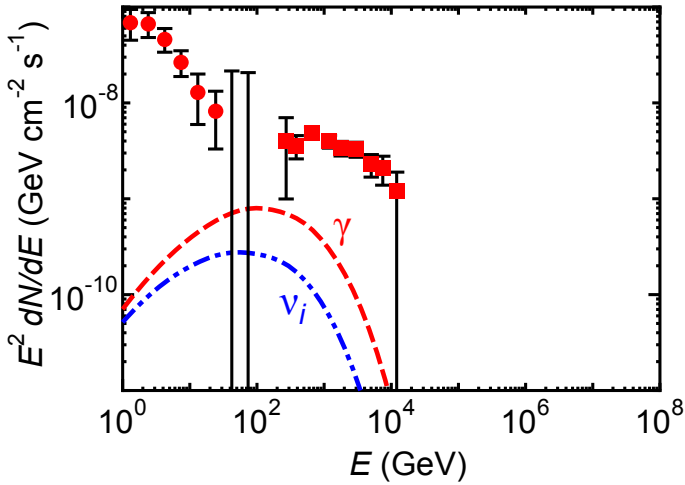} \caption{Predicted gamma-ray flux
(dashed line) and neutrino flux (two-dot dashed line) from the CMZ when
$\dot{m}=4.2\times 10^{-6}$ and $\lambda=0.01$. Filled circles and
squares are the Fermi and HESS observations, respectively
\cite{2013ApJ...762...33Y,2006Natur.439..695A}. \label{fig:now}}
\end{figure}

\begin{figure}
\includegraphics[width=90mm]{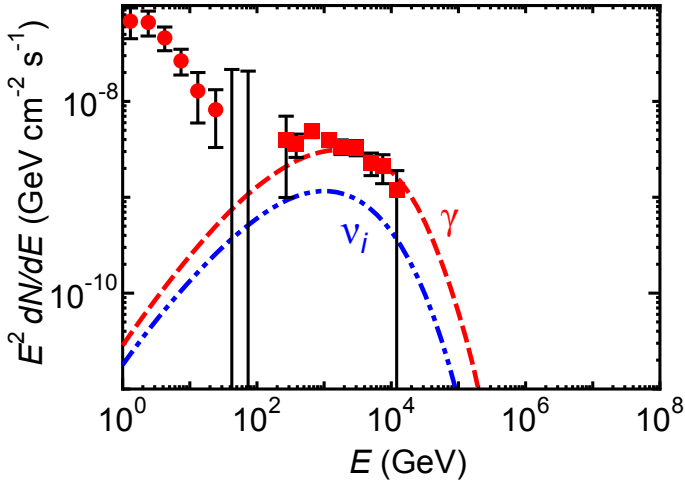} \caption{Same as Fig.~\ref{fig:now} but for
 $\dot{m}=0.001$, and $\lambda=5\times 10^{-4}$.\label{fig:HESS}}
\end{figure}

\begin{figure}
\includegraphics[width=90mm]{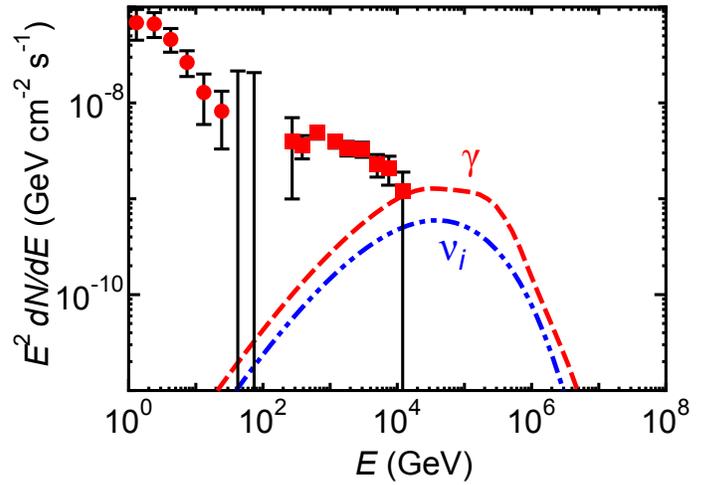} \caption{Same as Fig.~\ref{fig:now} but for
 $\dot{m}=0.001$, $\lambda=3\times 10^{-3}$, 
 $\zeta=0.18$, and $\eta_{\rm cr}=6\times 10^{-3}$.\label{fig:B1}}
\end{figure}

\begin{figure}
\includegraphics[width=90mm]{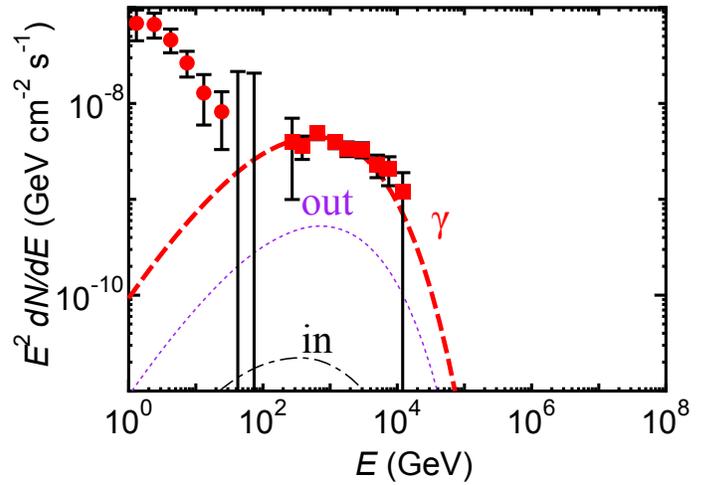} \caption{Same as Fig.~\ref{fig:now} but when
 $\dot{m}$ is variable (see text). Thick dashed line is the total
 gamma-ray flux. Thin dotted line (out) is the gamma-ray flux from
 $0.9\: R_{\rm CMZ} < r < R_{\rm CMZ}$. Thin dot-dashed line (in) is
 that from $0 < r < 0.1\: R_{\rm CMZ}$.\label{fig:var}}
\end{figure}

Figure~\ref{fig:now} shows the gamma-ray flux and neutrino flux (per
flavor) from the CMZ when $\dot{m}=4.2\times 10^{-6}$ and $\lambda=0.01$
regardless of $t$. Other parameters are the fiducial ones. For
comparison, we show the GeV and TeV gamma-ray fluxes obtained with Fermi
and HESS observations
\cite{2013ApJ...762...33Y,2006Natur.439..695A}. The predicted gamma-ray
flux is much smaller than the observations. The fraction of $\lambda\sim
0.5$ is needed in order that the flux is comparable to the observations
at $E\sim 1$~TeV. However, as is noted above, the value of $\lambda=0.5$
is probably too large for the actual CMZ. The gamma-ray and neutrino
spectra are similar and they are not represented by a power-law, because
they reflect the proton spectrum [Eq.~(\ref{eq:sp})].

Recent studies have indicated that the current activity of Sgr~A$^*$ is
exceptionally small, and that the average accretion rate more than $\sim
100$~yrs ago might be much larger and it could be as much as
$10^3$--$10^4$ times the current one
\cite{1996PASJ...48..249K,2000ApJ...534..283M,2006PASJ...58..965T,2013PASJ...65...33R}.
Thus, we calculate the gamma-ray and neutrino fluxes when
$\dot{m}=0.001$ and $\lambda=5\times 10^{-4}$ regardless of $t$. The
drop of activity in the past $\sim $100~yrs does not affect the results
because the diffusion time of CRs is much larger than 100~yrs. Other
parameters are the same as the fiducial ones. These give the typical
energy of $E_{p,\rm eq}=3.4\rm\: TeV$ from Eq.~(\ref{eq:geq}). Note that
for given $\dot{m}$ and $M_{\rm BH}$, any combinations of parameters
($\alpha$, $\zeta$, $\beta$, and $R_{\rm acc}/R_{\rm S}$) that give the
same $E_{p,\rm eq}$ give the same spectrum. Moreover, any combinations
of $\lambda$ and $\eta_{\rm cr}$ that give the same $\lambda\eta_{\rm
cr}$ give the same spectrum. Figure~\ref{fig:HESS} shows the results of
this model; the gamma-ray spectrum at $E\sim 0.2$--10~TeV well
reproduces that obtained with HESS. If observations for $E\gtrsim
10$~TeV become available in the future (e.g. Cherenkov Telescope Array;
CTA \cite{2013APh....43....3A}, High Altitude Water Cherenkov detector;
HAWC \cite{2013APh....50...26A}), our model predicts a soft gamma-ray
spectrum in that energy band. Since the designed sensitivities of CTA
and HAWC at $E\sim 10$--100~TeV are better than that of HESS, the flux
predicted in Fig.~\ref{fig:HESS} could be easily detected. The gamma-ray
image taken with HESS appears to coincide with the CMZ
\cite{2006Natur.439..695A}, which supports this model. Since the
apparent size of the CMZ is $\sim 3^\circ\times 0.5^\circ$, it can be
well resolved by CTA with a resolution of $\sim 1'$
\cite{2013APh....43....3A}. Detailed maps of gamma rays will reflect not
only the distribution of molecular gas but that of CRs. The latter may
reflect the history of Sgr A$^*$ activities, if the activities
significantly change on the diffusion time scale of the CRs.

From the current neutrino observations with IceCube, the flux of $\sim
3\times 10^{-8}\rm\: GeV\rm cm^{-2}\: s^{-1}$ can be attributed to that
from the Galactic center \cite{2014PhRvD..90b3010A}. Since the predicted
fluxes in Figs.~\ref{fig:now} and~\ref{fig:HESS} are smaller than that,
they are consistent with the observations. However, as the statistics of
neutrinos improve, we may detect an excess in the direction of the
Galactic center in the future. In particular, observations with KM3NeT
would be useful to detect the CMZ as a neutrino source if the flux is
$\gtrsim 10^{-9}\rm\: GeV\: cm^{-2}\:
s^{-1}$\cite{2013NIMPA.725...45S}. Our model predicts that the neutrino
image should coincide with the gamma-ray image because both are the
results of $pp$ interactions.

Since parameters for the RIAF have some uncertainties, we adopt another
model that can reproduce the neutrino flux at $\sim 1$~PeV obtained with
IceCube \cite{2015ApJ...806..159K}. We change the parameter for the
turbulence to $\zeta=0.18$, and the acceleration efficiency of CRs in
the RIAF to $\eta_{\rm cr}=6\times 10^{-3}$. We take $\dot{m}=0.001$,
which is the same as that in Fig.~\ref{fig:HESS}, and $\lambda=3\times
10^{-3}$ so that the flux at $E\sim 1$~TeV is consistent with the HESS
observations. Other parameters are the fiducial ones. For these
parameters, the typical energy is $E_{p,\rm eq}=160\rm\: TeV$
(Eq.~\ref{eq:geq}), which is much larger than that of the model in
Fig.~\ref{fig:HESS}. The results are shown in Fig.~\ref{fig:B1}.  If
this is the case, a relatively hard gamma-ray spectrum would be observed
by CTA at $\sim 10$--100~TeV, and could be discriminated from the
spectrum in Fig.~\ref{fig:HESS}. The gamma rays at $E\lesssim 1$~TeV
should have another origin.

So far we have assumed that the accretion rate, $\dot{m}$, is
constant. Here, we discuss the effects of variable $\dot{m}$. The x-ray
light curve of Sgr~A$^*$ in the past 500~yrs was derived in
Ref.~\cite{2013PASJ...65...33R}. They showed that the x-ray luminosity
is $L_X\sim 10^{39}\rm\: erg\: s^{-1}$ in the past 50--500 yrs, and then
it dropped to the current value of $L_X\sim 10^{33}$--$10^{35}\rm\:
erg\: s^{-1}$. The x-ray luminosity before 500~yrs ago is less
constrained. Upper limits are placed to down to about $8\times
10^{40}\rm\: erg\: s^{-1}$ for several periods within the past $4\times
10^4$~yrs \cite{2002A&A...389..252C,2014IAUS..303..333P}. Before that,
upper limits are $\sim 10^{41}$--$10^{42}\rm\: erg\: s^{-1}$
\cite{2002A&A...389..252C,2014IAUS..303..333P}.  The x-ray luminosity of
a RIAF is proportional to $\dot{m}^2$
\cite{2014ARA&A..52..529Y}. Assuming that the x-ray luminosity follows
the above observations and upper limits, and that $t_0=10^7$~yr is the
current time, we set $\dot{m}=0.03$ for $0<t<t_0-4\times 10^4\:{\rm
yr}$, $\dot{m}=0.01$ for $t_0-4\times 10^4\:{\rm yr}<t<t_0-1\times
10^4\:{\rm yr}$, $\dot{m}=0.001$ for $t_0-1\times 10^4\:{\rm
yr}<t<t_0-50\:{\rm yr}$, and $\dot{m}=4.2\times 10^{-6}$ for
$t_0-50\:{\rm yr}<t<t_0$. Other parameters, including the fiducial
values, are time-independent, except for $\lambda=4\times 10^{-5}$ and
$\zeta=0.025$, which are chosen to be consistent with observations at
$E\sim 0.2$--10~TeV.

Figure~\ref{fig:var} shows the results. The gamma-ray flux from the
outer region originates in CRs injected in earlier times. Since
$\dot{m}$ decreases as time advances, the typical energy of the CRs,
$E_{p,\rm eq}$, in the CMZ should decrease from the outer region to the
inner region [Eq.~(\ref{eq:geq})]. We obtain $E_{p,\rm eq}=2.3$~TeV,
when $\dot{m}=0.03$. However, the shape of gamma-ray spectrum from $0 <
r < 0.1\: R_{\rm CMZ}$ is not much different than that from $0.9\:
R_{\rm CMZ} < r < R_{\rm CMZ}$. The peak gamma-ray energy of the former
is only a factor of 2 smaller than the latter. Most of the gamma-ray
flux from the CMZ is associated with CRs injected when $\dot{m}=0.03$,
because they are injected during most of the past diffusion time
($t_0-t_{\rm diff,C}<t<t_0-4\times 10^4$~yr), where $t_{\rm
diff,C}=R_{\rm CMZ}^2/(6\kappa)\sim 1.6\times 10^5$~yr (at $E_p\sim
1$~TeV). Those CRs prevail in the CMZ even at present. CRs of $E\lesssim
1$~TeV injected when $\dot{m}\leq 0.001$ are located at $r\lesssim
0.25\: R_{\rm CMZ}$. However, their short injection time scale compared
with $t_{\rm diff,C}$ and the smaller injection rate ($L_{p,\rm
tot}\propto \dot{m}$) make their contribution to the gamma-ray flux
smaller.  In other words, the contribution of CRs is represented in the
form of $\int \lambda L_{p,\rm tot}dt$ integrated for the past diffusion
time. Moreover, since the higher-energy CRs ($\gg 1$~TeV) have shorter
diffusion times, they escape faster from the CMZ than lower-energy
CRs. This also makes the gamma-ray spectrum from the older CRs
softer. Of course, if the typical CR energy, $E_{p,\rm eq}$,
significantly varies in the past, while $\dot{m}$ does not much decrease
[see (Eq.~\ref{eq:geq})], the gamma-ray spectrum can change across the
CMZ.  Note that the gamma-ray luminosity of the RIAF is proportional to
$\dot{m}^2$ and it is $\sim 0.001\: L_{p, \rm tot}\sim 10^{37}\rm erg\:
s^{-1}$ when $\dot{m}=0.001$ \cite{2015ApJ...806..159K}. Although this
is larger than the current total gamma-ray luminosity of the CMZ at
$\sim 1$~TeV ($\sim 4\times 10^{34}\rm erg\: s^{-1}$), the gamma-rays
from the RIAF cannot be observed at present. This is because the
gamma-ray luminosity of the RIAF almost immediately changes with
$\dot{m}$, owing to the short ($<1$~yr) diffusion or escape time of the
CRs in the RIAF \cite{2015ApJ...806..159K}.

\section{V. gamma rays from the CMZ around Centaurus~A}

Since some LLAGNs other than Sgr~A$^*$ also have their own CMZs, gamma
rays and neutrinos may be created there. As for neutrinos from their
RIAFs, LLAGNs with $\dot{m}\sim 0.01$--0.1 most contribute to the
neutrino flux on the Earth, because the neutrino production rate in a
RIAF is sensitive to $\dot{m}$ and RIAFs are realized when
$\dot{m}\lesssim 0.01$--0.1 \cite{2015ApJ...806..159K}. If a fraction
$\lambda\lesssim 10^{-3}$ of the CR protons accelerated in those RIAFs
enter their CMZs as is the case of Sgr~A$^*$, the production rate of
neutrinos in the CMZs is smaller than that in the RIAFs. Thus the
contribution of the former to the overall neutrino flux on the Earth is
expected to be smaller than the latter. However, if the RIAF in a nearby
galaxy is well covered with massive molecular gas or a CMZ and $\dot{m}$
is relatively large, the gamma rays from the CMZ may still be detectable
as an individual source.

\begin{figure}
\includegraphics[width=90mm]{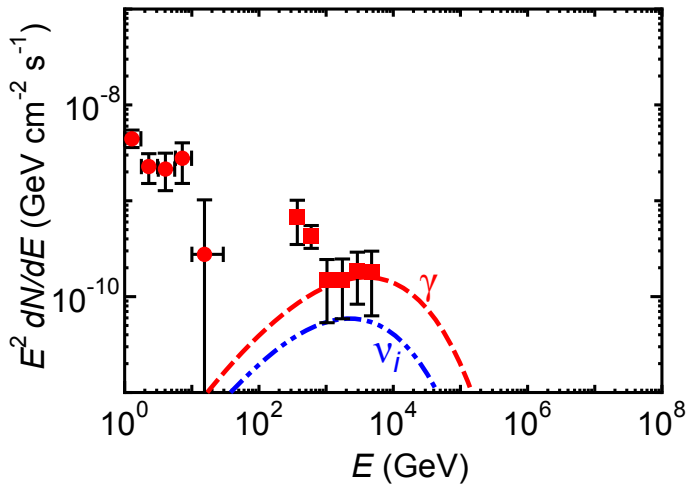} \caption{Predicated gamma-ray flux (dashed
 line) and neutrino flux (two-dot dashed line) from Cen~A.  Parameters
 are shown in the text. Filled circles and squares are the Fermi and
 HESS observations, respectively
 \cite{2010ApJ...719.1433A,2009ApJ...695L..40A}.\label{fig:CenA}}
\end{figure}

In Fig.~\ref{fig:CenA}, we show as an example the gamma-ray flux from
Centaurus~A (Cen~A), which is a nearby radio galaxy and for which the
origin of the gamma rays is under debate
\cite{2012PhRvD..85d3012S,2013ApJ...770L...6S}. The distance to Cen~A is
assumed to be 3.84~Mpc. We do not include the absorption of the gamma
rays. The radius, thickness, and mass of the CMZ are $R_{\rm
CMZ,obs}=$195~pc, $H_{\rm CMZ,obs}=$195~pc, and $M_{\rm CMZ}=8.4\times
10^7\: M_\odot$, respectively \cite{2014A&A...562A..96I}. The mass of
the SMBH is $M_{\rm BH}=5\times 10^7\: M_\odot$
\citep{2009MNRAS.394..660C}. We choose $\dot{m}=0.01$, $\lambda=0.02$,
and $\zeta=0.03$ in order to reproduce the HESS results. Other
parameters are the same as the fiducial ones. These give the typical
energy of $E_{p,\rm eq}=7.9\rm\: TeV$ from Eq.~(\ref{eq:geq}). Cen~A has
a prominent cold gas disc \cite{1997SSRv...81....1H} and, thus, the
effective covering factor $\lambda$ may be larger than that of
Sgr~A$^*$. Figure~\ref{fig:CenA} shows that our model can reproduce the
HESS observations at $E\gtrsim 1$~TeV, although another component is
required at $E\lesssim 1$~TeV. We note that the actual gamma-ray flux
from the CMZ could be smaller, if that from the CRs accelerated by other
mechanisms (e.g. acceleration in the electronic field at the base of
jets \cite{2014Sci...346.1080A}) cannot be ignored. In fact, jets have
been observed in Cen~A \cite{2007ApJ...670L..81H}.

\section{VI. Summary}

We have shown that TeV gamma rays from the Galactic center can be used
to test a model in which low-luminosity active galactic nuclei (LLAGNs)
are the source of neutrinos observed with IceCube. In this model,
protons are accelerated in the radiatively inefficient accretion flows
(RIAFs) in the LLAGNs, and neutrinos are created through $pp$ and
$p\gamma$ interactions in the flows. Since Sgr~A$^*$ at the center of
the Galaxy is an LLAGN, we expect that the protons are being accelerated
in Sgr~A$^*$ and injected into the interstellar space. 

In this study, we found that the central molecular zone (CMZ)
surrounding Sgr~A$^*$ works as an effective target of the high-energy
protons escaped from the RIAF and gamma rays and neutrinos are created
there through $pp$ interactions. We showed that our model can explain
the gamma rays observed by HESS at $E\sim 0.2$--10~TeV, if the accretion
rate on Sgr~A$^*$ was $\sim 10^3$ times larger in the past than it is
today as indicated by previous studies, and if the typical energy of the
CR protons is $\sim$~TeV. In the near future, CTA could observe gamma
rays at $\sim 10$--100~TeV, which could be used to estimate the typical
energy of the CR protons more precisely. The gamma-ray emission from
some nearby galaxies could be attributed to this mechanism if their
LLAGNs are surrounded by molecular gas. Future comparison between
observed gamma-ray and neutrino spectra and images would be useful to
confirm this model.

\begin{acknowledgments}
 This work is partly supported by the challenge support program of Osaka
 University and by JSPS KAKENHI Grant Number 15K05080 (Y.F). This work
 is also partly supported by Grant-in-Aid for JSPS Fellowships
 No. 251784 (S.S.K.). K.M. acknowledges Institute for Advanced Study for
 continuous support.
\end{acknowledgments}

%

\end{document}